\documentclass[%
reprint,
    amsmath,
    amssymb, 
 aps,
]{revtex4-2}

\usepackage{graphicx}
\usepackage{dcolumn}
\usepackage{bm}

\begin{document}

%-------------------------------------------------------
%--------------------title------------------------------
%-------------------------------------------------------
\title{Effect of microstructure on fatigue properties of hyperelastic materials}

\author{Anna Stepashkina}%
\author{Fuguang Chen}
\author{Lipeng Chen}
 \email{chenlp@zhejianglab.com}
\affiliation{%
Research Center for New Materials Computing, Zhejiang Lab, Hangzhou, China
}%

%\date{\today}

%-------------------------------------------------------
%--------------------------abstract---------------------
%-------------------------------------------------------

\begin{abstract}
Homogenization is a technique for the analysis of complex materials by replacing them with equivalent homogeneous materials that exhibit similar properties. By constructing a three-dimensional (3D) porous material model and employing homogenization technique, effective properties of the hydrogel pore structure were calculated. It is found that the microstructure of hyperelastic materials has considerable influence on their macroscopic mechanical properties, pores with a radius of up to 65 microns at a small strain can significantly reinforce material structure and improve its fatigue resistance. This work highlights the potential of engineering pore structures for the enhancement of mchanical properties and durability of hydrogels.
\end{abstract}

%-------------------------------------------------------
%----------------------introduction---------------------
%-------------------------------------------------------
\maketitle
\section{Introduction}
Due to their unique elastic properties that remain consistent regardless of the rate of deformation, hyperelastic materials are widely used in the production of biological materials, foams, rubbers \cite{wang2023simulating, baur2023porous, zhong2011fabrication, shahzad2015mechanical, khaniki2023hyperelastic, fkas2024stiffness, sutarya2021effect, pan2024silk}. Some hyperelastic materials can be compressed by a strain of more than 70\% \cite{khaniki2023hyperelastic, sutarya2021effect, treloar1943elasticity}. Additionally, these materials often exhibit a Poisson ratio close to 0.5. When analyzing mechanical properties of hyperelatsic materials using classical mechanics models, such as Cauchy's theorem, one typically obtains incorrect results. Hyperelastic materials can be viewed as a special case of Cauchy elastic materials. In order to accurately describe  deformation properties of hyperelastic materials and calculate the stress-strain curve, rheological mechanics models are deemed the most appropriate ones. These hyperelastic rheological models work as types of constitutive models for ideal elastic materials, with the stress-strain relation derived from a strain energy density function

\begin{equation}
\sigma = J^{-1} \frac{ \delta{W_s} }{ \delta {F} } F^T.
\end{equation}

Here, $J$ is the elastic volume ratio ($J = \mathrm{det} F$), $F$ is the deformation gradient, $W_{s}$ is the strain energy density. The stress-strain relation for hyperelastic materials is described by nonlinear mechanical models, including nonlinear elastic and nonlinear viscoelastic models \cite{kim2012comparison, bergstrom2015mechanics}. The most famous models are neo-Hookean \cite{kim2012comparison, treloar1943elasticity,  pence2015compressible}, Mooney-Rivlin \cite{kim2012comparison, melly2021review}, Ogden \cite{kim2012comparison, melly2021review, ogden2004fitting}, Storakers \cite{storaakers1986material}.

The neo-Hookean model \cite{kim2012comparison, treloar1943elasticity,  pence2015compressible} is a hyperelastic material model, which can be viewed as an extension of Hooke's law that can predict the nonlinear stress-strain behavior of materials undergoing large deformations. In contrast to linear elastic materials, the stress-strain relation of a neo-Hookean material is initially linear and goes to plateau at certain strains. In neo-Hookean model, one does not account for the dissipation of energy as heat when straining the material, and perfect elasticity is assumed at all stages of deformation.

Based on the statistical thermodynamics of cross-linked polymer chains, the neo-Hookean model is usable for plastics and rubber-like substances \cite{kim2012comparison, treloar1943elasticity,  pence2015compressible}. Since polymer chains can move relative to each other when a stress is applied, cross-linked polymers will behave in a neo-Hookean manner. However, polymer chains will be stretched to a maximum point at which covalent cross links will allow, leading to a dramatic increase in the elastic modulus of the material. For neo-Hookean model, the energy density $W_{s}$ is defined as 

\begin{equation}
W_s=\frac{\mu}{2}(I-3)-\mu\mathrm{ln}J+\frac{\lambda}{2}\mathrm{ln}^2J,
\end{equation}

where $I$ is the deformation tensor trace, and $\mu$, $\lambda$ are Lame parameters,

\begin{equation}
\mu=\frac{E}{2(1+\nu)},
\end{equation}
\begin{equation}
\lambda=\frac{E\nu}{(1+\nu)(1-2\nu)}.
\end{equation}

Here, $\nu$ is the Poisson ratio, and $E$ is the  Young's modulus.

The neo-Hookean model has been superseded by the Mooney-Rivlin model \cite{kim2012comparison}, which can describe the stress-strain relation with high accuracy. It works well from small to medium strains \cite{melly2021review}. The Mooney-Rivlin model has special phenomenological parameters. For the one direction compression, $W_s$ can be written as

\begin{equation}
W_s=C_{10}(I_1-3)+C_{01}(I_2-3)+\frac{\kappa}{2}(J-1)^2.
\end{equation}

Here, $\kappa =\frac{E}{3(1-2\nu)}$ is the bulk modulus, $C_{10}$, $C_{01}$ are Mooney-Rivlin parameters. These parameters can be obtained by fitting experimental test data with the analytical expression describing the stress-strain relation. $C_{10}$, $C_{01}$ can be calculated from the system of equations

\begin{equation}
\sigma_{i} = 2(C_{10}+\frac{C_{01}}{\epsilon_{i}})(\epsilon_{i}-\frac{1}{\epsilon_{i}^2}).  
\end{equation}

Here, $\sigma_{i}$ is the stress corresponding to the deformation $\epsilon_{i}$. 

In the Ogden model \cite{kim2012comparison, melly2021review, ogden2004fitting}, the strain energy function $W_s$ is defined as 

\begin{equation}
W_{s}= \frac{\mu_{\mathrm{mod}}}{\alpha}( \lambda_1^\alpha+\lambda_2^\alpha+\lambda_3^\alpha-3) + \frac{\mu_{\mathrm{mod}}}{2\alpha\beta}(J^{-\alpha\beta}-1)^2,
\end{equation}

with 

\begin{equation}
\beta = \frac{\nu}{1-2\nu},  
\end{equation}
\begin{equation}
\mu_{\mathrm{mod}} = 0.5 \frac{E}{1+\nu},
\end{equation}

where $\mu_{\mathrm{mod}}$ is the shear modulus, $\alpha$ is a constant which can be determined from the equation 

\begin{equation}
\sigma_{i}=2\frac{\mu_{\mathrm{mod}}}{\alpha}(\lambda^\alpha_{i}-(\frac{1}{\sqrt{\lambda_{i}}})^\alpha).    \end{equation}

The Mooney-Rivlin model and Odgen model work well only from small to medium deformation, for the large deformation one needs to employ the Storakers model \cite{storaakers1986material}, which yields better results in the description of stress-strain relation. The Storakers model is a modern version of previous phenomenological models  \cite{kim2012comparison, bergstrom2015mechanics, treloar1943elasticity,  pence2015compressible, melly2021review, ogden2004fitting}, often used to model highly compressible samples. Assuming that only elastic deformation is allowed, one can obtain $W_s$ of an isotropic hyperelastic material   

\begin{equation}
W_s=\frac{2\mu_{\mathrm{mod}}}{\alpha}(\lambda_1^{\alpha}+\lambda_2^{\alpha}+\lambda_3^{\alpha}-3+\frac{1}{\beta}(J^{-\alpha\beta}-1)). 
\end{equation}

Determining empirical parameters is indeed crucial when using rheological models to describe the mechanical properties of materials. These parameters enable us to construct stress-strain curves and analyze various applied problems, such as fatigue behavior. However, in traditional rheological models, one often neglects the microstructure of materials, which largely restricts their application to materials with complex microstructures, such as porous hyperelastic materials. 

In this work, we proposed a novel approach using the homogenization technique to predict the mechanical properties of porous hyperelastic materials. Homogenized properties are used to model the mechanical behavior of composite materials or heterogeneous structures. This significantly simplifies the analysis and design processes while capturing the overall mechanical behavior of materials. By including microstructures into the model, it is possible to improve the predictive accuracy and develop more realistic rheological models for hyperelatic materials.

%-------------------------------------------------------
%----------------------homogenisation-------------------
%-------------------------------------------------------

\section{Mechanical properties of homogenized hyperelastic materials}

Experimental studies of mechanical properties and microstructures of hyperelastic materials, in particular, hydrogels and biological materials, have shown that these materials have porous structures with pore sizes ranging from 50 um to 100 um \cite{baur2023porous, zhong2011fabrication, khaniki2023hyperelastic, pan2024silk, campbell2017development}. Following the research in Ref.~\cite{pan2024silk}, we employ aforementioned methods to construct approximated models to calculate the stress-strain curve of hydrogel. To this end, we need to obtain the empirical parameters in the model, which can be determined by fitting the model to a set of measurements. The method for determining  empirical parameters depends on the quality of the experiment, the range of deformations, and the loading method. To find the accurate solution, it is often necessary to carry out a large number of measurements.

\begin{figure}
    \includegraphics[width=0.48\textwidth]{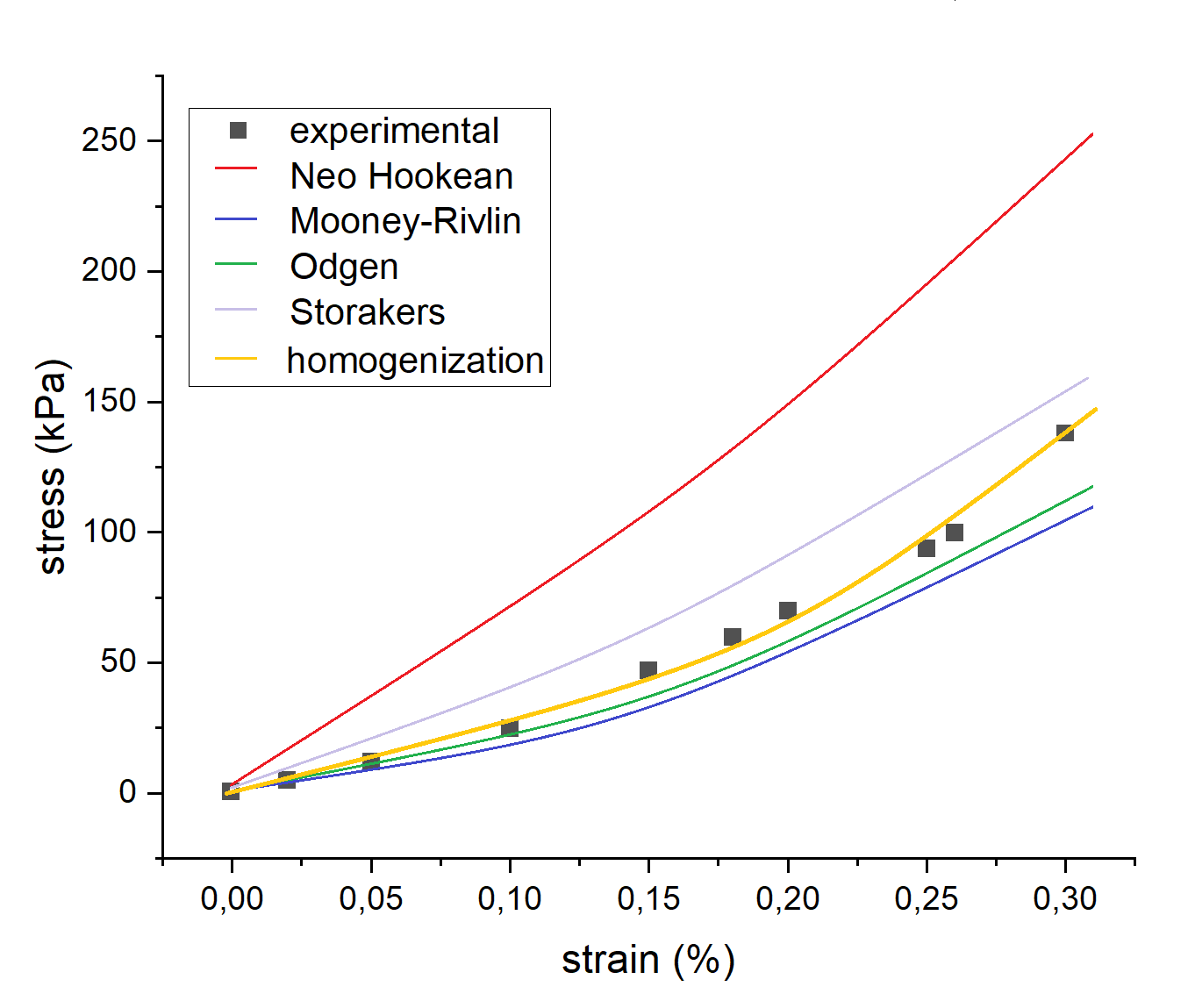}
    \caption{the stress-strain curve for hydrogel}
    \label{fig1}
\end{figure}

For the neo-Hookean model, we can build the dependence directly since all parameters are known. The optimal values of empirical parameters in other models (the Mooney-Rivlin model, the Storakers model, the Odgen model) are determined using the least squares method based on the experimental stress-strain curve. Fig.~\ref{fig1} shows the stress-strain curve of hydrogel calculated by neo-Hookean, Mooney-Rivlin, Odgen and Storakers models. These rheological models are often based on certain simplifications and may not capture all the complexities of material behavior, especially in the presence of pores or other microstructural features. To accurately describe the effect of pores/inclusions on mechanical properties, we build a homogenized model of the material using the finite element method. Homogenization refers to the process of calculating the effective properties of a material, which involves treating the porous structure as a whole material.  

\begin{figure}
    \includegraphics[width=0.48\textwidth]{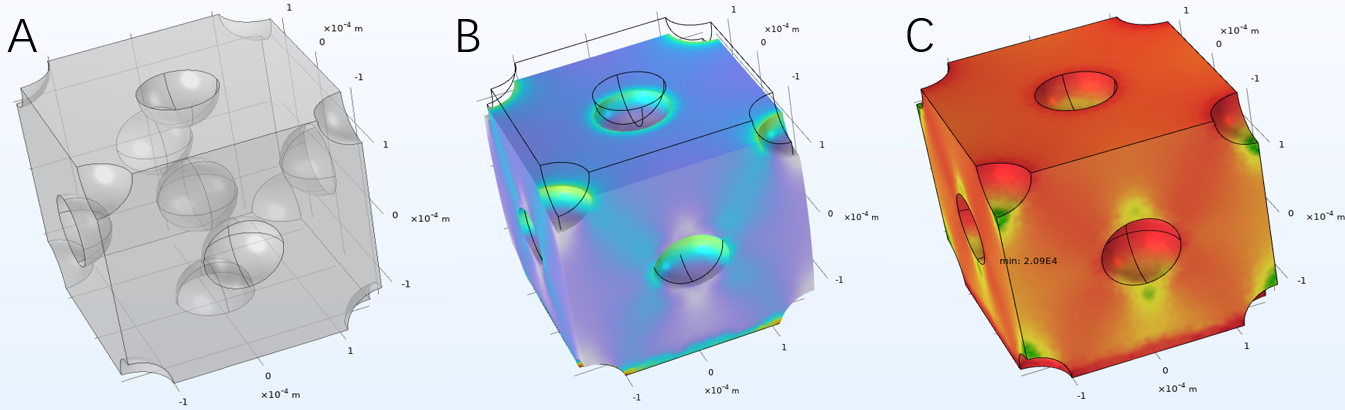}
    \caption{A: microstructure model of hydrogel (pores volume fraction 10\%); B: deformation during pressing (pores volume fraction 10\%, strain 20\%), C: fatigue (pores volume fraction 10\%, strain 20\%)}
    \label{fig2}
\end{figure}

Constructing the microstructure of a material is a complex process that requires careful design. The arrangement of pores in the material is chaotic, and their shapes exhibit complex geometric structures \cite{baur2023porous, zhong2011fabrication, pan2024silk}. In the homogenization process, a cube with pores was considered as a periodic element of the material. To a first approximation, the pores were spherical objects located periodically,

\begin{itemize}
    \item one sphere in the center of the cube, 
\end{itemize}
\begin{itemize}
    \item 6 spheres in the center of the cube faces, 
\end{itemize}
\begin{itemize}
    \item 8 spheres at the vertices of the cube. 
\end{itemize}
The cell structure is shown in Fig.~\ref{fig2} (A). The size of the cubic cell is 300 um, and the diameter of the sphere is calculated based on the pores volume fraction 10\% of the material (radius r $\approx$ 47 um).

Calculating homogenized properties involves the simulation of averaged properties of the material within a small volume that exhibits structural periodicity. Instead of simulating the entire material, a small representative volume element (RVE) is selected. The RVE is chosen to be small enough to capture the local microstructure but large enough to represent the overall behavior of the material. Homogenized properties were averaged by the properties of each constituent phase within the RVE, which can be determined by the system of equation with periodical boundary conditions, 
\begin{equation}\label{SOE}
0=\nabla(F\cdot{S})^T+\textbf{F}\cdot{v}
\end{equation},
\begin{equation}
F=I+\nabla\textbf{u}. 
\end{equation}
Here, $F$ is the deformation gradient tensor, which describes how the body deforms in response to applied forces or displacements. It relates the current configuration of the body to its reference (undeformed) configuration. $S$ is the second Piola-Kirchhoff stress tensor, which represents the stress state within the body. It accounts for the internal forces within the material. $\textbf{F}\cdot{v}$ represents the external body forces per unit volume acting on the body. These forces include gravity, pressure, or other applied loads. $\nabla\textbf{u}$ is the displacement gradient tensor, and $I$ is the stretch tensor which represents the relative deformation or change in shape of the body. Eq.~\ref{SOE} represents the equilibrium condition for a deformable body, which states that the sum of internal and external forces acting on a small volume element of the body is zero.

Upon solving the homogenization problem, one obtains Poisson's ratio, Young's modulus and shear modulus for the porous hydrogel. The stress-strain curve obtained by the homogenization method is displayed in Fig.~\ref{fig1}, which increases accuracy significantly as compared to empirical models. 
%-------------------------------------------------------
%----------------------fatigue--------------------------
%-------------------------------------------------------

\section{Fatigue analysis for homogenized material}
Since the homogenization method yields results in good agreement with experiment, we will systematically investigate how pores affect mechanical properties of the material. As a control simulation, we will study the fatigue of the material, which depends on factors such as the applied load, microstructure, external conditions. We consider a closed system at temperature 20°C and pressure 105 Pa. During operation, hydrogel materials are usually subjected to cyclic loading, and we consider 4 compression modes, 
\begin{itemize}
    \item 0\%-5\%-0\%,
\end{itemize}
\begin{itemize}
    \item 0\%-10\%-0\%,
\end{itemize}
\begin{itemize}
    \item 0\%-20\%-0\%,
\end{itemize}
\begin{itemize}
    \item 0\%-30\%-0\%. 
\end{itemize}

The higher the deformation, the faster fatigue develops, but cycles to failure can change nonlinearly. The microstructures, such as pore size and volume fraction can affect significantly the fatigue characteristics of the material. The coarse-grained structure increases fatigue, and reaching a certain concentration of inclusions can increase cycles to failure. As a crucial parameter for studying the influence of inclusion size, we will take the volume fraction of pores, according to which the diameter of spherical inclusions is calculated. We study the volume fraction of pores ranging from 
5\% to 35\% (sphere diameter from $\approx$ 37 um to 72 um). The fatigue model is sketched in Figs~\ref{fig2}B, C.

\begin{figure}
    \includegraphics[width=0.48\textwidth]{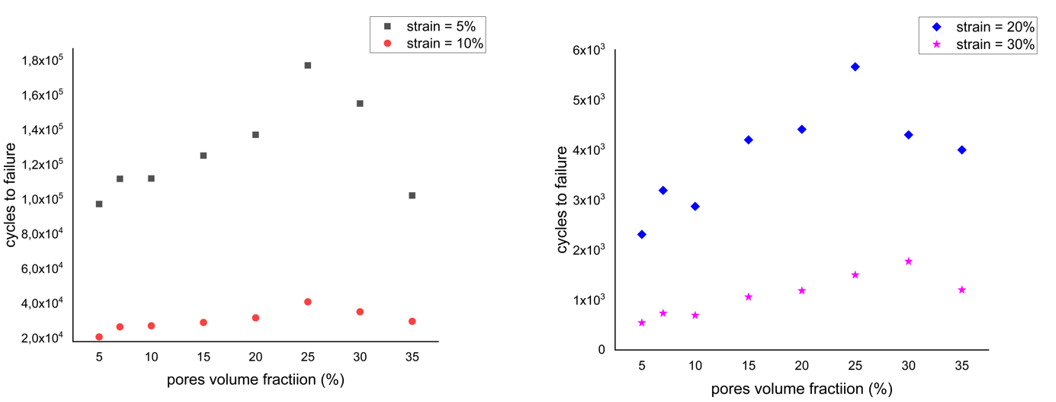}
    \caption{Fatigue for different pores volume fractions.}
    \label{fig3}
\end{figure}

The highest mechanical stress concentrations are observed at the boundaries of spherical inclusions within the hyperelastic material, as illustrated in Fig~\ref{fig2}B. This phenomenon can be attributed to the abrupt changes in material properties at the interface between the inclusions and the surrounding matrix. When an external load is applied to the material, the stress tends to concentrate at the boundaries of the inclusions because of the mismatch in stiffness or elasticity between two phases. The stress builds up at the interfaces, leading to localized regions of high stress. This so-called stress concentration effect has important implications on 
the mechanical behavior and failure mechanism of the material. The regions with high stress concentrations are more susceptible to the initiation and propagation of defects, such as cracks or voids. These defects can act as stress concentrators, further exacerbating the stress concentrations and potentially leading to premature failure of the material. 

The number of cycles to failure means the minimum value of cycles when irreversible deformations occur. It leads to small defects, and further exploitation of the material can result in further destruction, even total destruction. By treating the hyperelatic material as a composite, the pores can act as reinforcements, similar to the role of fibers in a fiber-reinforced composite. The pores can hinder the propagation of cracks and defects, leading to improved toughness and resistance to failure. In addition, the pores can also facilitate the energy absorption and dissipation, which contributes to the overall mechanical performance of the material.

Fig.~\ref{fig3} displays the fatigue behavior of hydrogel as a function of pores volume fraction at strains of 5\%, 10\%, 20\%, 30\%. Under cyclic loading with a small compressive strain of 5\%, the number of compression cycles increases with the pore volume fraction, leading to the reinforced structure. Upon further increasing the pore volume fraction ($\geq$ 25\%, pore diameter r $\geq$ 65 um), the fatigue characteristics of the material deteriorate sharply. For cyclic loadings with a large value of compressive strain, one can also observe the reinforced effect of pores, but not so significantly. It is found that the minimum number of cycles at a compressive strain of 30\% for a maximum pore volume of 35\% (r $\approx$ 37 um) is 2190.

%-------------------------------------------------------
%----------------------conclusion-----------------------
%-------------------------------------------------------

\section{Conclusion}
By using the homogenization technique and building a 3D porous material model, we derived effective properties of porous hydrogels. Homogenization involves treating the complex microstructure as an equivalent homogeneous material, allowing for simpler analysis while still capturing the overall behavior. This approach enables us to calculate the effective elastic modulus, Poisson's ratio, and other mechanical properties of the hydrogel with pores. By modelling the deformation and stress distribution within the hydrogel under various loading conditions, we obtain highly accurate stress-strain curve. Under cyclic loading, we also studied the fatigue behavior of the porous hydrogel, which provides detailed information on the durability and lifespan of the hydrogel under realistic service conditions. It is found that pores with a radius of up to 65 um at a small compressive strain can significantly strengthen the material structure and improve the fatigue performance.

The combination of homogenization technique, a detailed 3D microstructure model  and fatigue analysis offers a comprehensive understanding of the mechanical response of hyperelastic materials like hydrogels. This knowledge have important applications in fields such as biomedical engineering, soft robotics, and tissue engineering, where mechanical properties and durability of these materials play a critical role in their functionality and reliability. Work in this direction is in progress.

%-------------------------------------------------------
%----------------------references-----------------------
%-------------------------------------------------------

\bibliography{bibl} 
\bibliographystyle{unsrt}

\end{document}